\documentclass[12pt]{article}
\usepackage{epsfig}

\newlength{\oline}

\newcommand{\ol}{\vspace{\oline}}
\begin{document}
\begin{center}
{\large \bf PHOTON - AXION CONVERSION CROSS SECTIONS IN A RESONANT
CAVITY\\} \ol
{\bf Dang Van Soa$^{(a)}$}, {\bf Hoang Ngoc Long$^{(b)}$}\\[0.5cm]
{\bf and}\\
{\bf Le Nhu Thuc$^{(a)}$}\\
(a): {\it Department of Physics, Hanoi University of Education,
Vietnam}\\
{\it E-mail:} dvsoa@assoc.iop.vast.ac.vn\\[0.5cm]
(b): {\it Institute of Physics, VAST, P. O.  Box 429, Bo Ho,
Hanoi 10000, Vietnam}\\
{\it Email:} hnlong@iop.vast.ac.vn\\
 \vspace{1cm}
{\bf Abstract}\\
\end{center}
\hspace*{0.5cm}Photon - axion conversions  in the resonant cavity
with the lowest mode are considered in detail by the Feynman
diagram method. The differential cross sections are presented and
numerical evaluations are given. It is shown that there is a
resonant conversion for the considered process, in which the
conversion cross sections are much larger than those of the wave
guide in the same conditions. Some estimates for experimental
conditions are given from our results.\\[0.2cm]
\begin{center}
\end{center}
\newpage
{\bf 1. Introduction}\\

\hspace*{0.5cm} The most attractive candidate for solution of the
strong-CP problem is  Peccei and Quinn mechanism, where the
CP-violating phase $\theta$ is explained by the existence of a new
pseudo-scalar field, called the axion~\cite{pec,wei,wil}. At
present, the axion mass is constrained by laboratory
searches~\cite{kim,pec1} and by astrophysical and cosmological
considerations~\cite{tur,tur1} to between $10^{-6} eV $ and
$10^{-3} eV$. If the axion has a mass near the low limit of order
$10^{-5} eV $, it is good candidate for the dark matter of the
universe. Besides that, an axino (the fermionic partner of the
axion) naturally appears in SUSY models~\cite{kim1,raj}, which
acquire a mass from three - loop Feynman diagrams in a typical
range between a few eV up to a maximum of ~$1$ keV
~\cite{vys,mas}. The candidates for dark matter can be appeared in
different models. They were done in multi - Higgs extension of the
standard model, in the 3-3-1 models
~\cite{ale,ale1,ale2} or in the supersymmetric and superstring theories~\cite{babu}.\\
\hspace*{0.5cm}Neutral pions, gravitons, hypothetical axions, or
other particles with a two-photon interaction can transform into
photons in external electric or magnetic fields,  an effect first
discussed by Primakoff ~\cite{pri}. This effect is the basis of
Sikivie's methods for the detection of axions in a resonant
cavity~\cite{sik}. He suggested that this method can be used to
detect the hypothetical galactic axion flux that would exist if
axions were the dark matter of the Universe. Various terrestrial
experiments to detect invisible axions by making use of their
coupling to photons have been proposed~\cite{van,long,dvsoa}, and
results of such experiments appeared recently~\cite{hag,mor,min}
.The experiment CAST~\cite{ira} at CERN searches for axions from
the sun or other sources in the universe. The experiment uses a
large magnet from LHC and searches for the conversion of axions
into photons. The potential of the CAST experiment for exotic
particles was discussed~\cite{ira1}.
 The purpose of this paper is to
consider conversions of photons into axions in the periodic EM
field of the resonant cavity. We show that there is a resonant
conversion for the considered process at the low energies. The
text is organized as follows. In Sec.II we give the matrix element
of the considered process. The differential cross section (DCS)
for conversions in the resonant cavity with the $ TM_{110}$ mode
is presented in Sec.III. Some estimates for experimental
conditions are given in the last section - Sec. VI.\\
{\bf 2. Matrix element}\\
\hspace*{0.5cm}The axion mass and its couplings to ordinary
particles are all inversely proportional to the magnitude $v$ of
the vacuum expectation value that spontaneously breaks the
$U_{PQ}(1)$ quasisymmetry which was postulated by Peccei and Quinn
and of which the axion is the pseudo-Nambu - Goldstone boson. For
the axion- photon system a suitable Lagrangian density is given
by~\cite{tur,kim1}

\begin{equation}
L = -\frac{1}{4}F_{\mu\nu}F^{\mu\nu} + g_{\gamma}\frac{\alpha}{4\pi}
\frac{\phi_a}{f_a}F_{\mu\nu}\tilde{F}^{\mu\nu} + \frac{1}{2}\partial_{\mu}\phi_a
\partial^{\mu}\phi_a - \frac{1}{2} m_a^2 \phi_a^2[1+{\cal O}(\phi_a^2/v^2)]
\end{equation}
where $\phi_a$ is the axion field, $m_a$ is its mass,
$\tilde{F}_{\mu\nu}=\frac{1}{2}\varepsilon_{\mu\nu\rho\sigma}F^{\rho\sigma}$,
and $f_a$ is the axion decay constant and is defined in terms  of
the axion mass $m_a$ by [9,12]: $f_a = f_{\pi}m_{\pi}\sqrt{m_u
m_d}[m_a(m_u + m_d)]^{-1}$. Interaction of axions to the photons
arises from the triangle loop diagram, in which two vertices are
interactions of the photon to electrically charged fermion and an
another vertex is coupling of the axion with fermion. This
coupling is model dependent and is given by: $ g_{\gamma}=
\frac{1}{2}\left(\frac{N_e}{N} - \frac{5}{3}- \frac{m_d - m_u}
{m_d + m_u}\right)$ where $ N = Tr(Q_{PQ} Q^2_{color})$ and $ N_e
= Q_{PQ} Q^2_{em}$. Tr represents the sum over all left - handed
Weyl fermions. $Q_{PQ}, Q_{em}$, and $Q_{color}$ are respectively
the Peccei- Quinn charge, the electric charge, and one of the
generators of $SU(3)_c$. In particular in the Dine- Fischler -
Srednicki - Zhitnitskii model ~\cite{din,din1}:
$g_{\gamma}$(DFSZ)$\simeq 0.36$, and in the Kim- Shifman-
Vainshtein- Zakharov model ~\cite{kim2,kim3}( where the axions do
not couple to light quarks and  leptons) :
$g_{\gamma}$(KSVZ)$\simeq -0.97$.\\
\hspace*{0.5cm}Consider the conversion of the photon $\gamma$ with
momentum $q$ into the axion $a$ with momentum $p$ in an external
electromagnetic field. For the above mentioned process, the
relevant coupling is the second term in (1). Using the Feynman
rules we get the following expression for the matrix
element~\cite{long}
\begin{equation}
\langle p|{\cal M}|q\rangle = -\frac{g_{a\gamma}}{2(2\pi)^2
\sqrt{q_0p_0}}\varepsilon_{\mu}(\vec{q},\sigma)\varepsilon^{\mu\nu\alpha\beta}q_{\nu}
\int_{V} e^{i\vec{k}\vec{r}}F_{\alpha\beta}^{class}d\vec{r}
\end{equation}
where $ \vec{k}\equiv \vec{q}-\vec{p}$ is  the momentum transfer
to the EM field, $g_{a\gamma}\equiv g_{\gamma}\frac{\alpha}{\pi
f_a}$= $g_{\gamma}\alpha m_a(m_u+m_d)(\pi
f_{\pi}m_{\pi}\sqrt{m_um_d})^{-1}$ and
$\varepsilon^{\mu}(\vec{q},\sigma)$ represents the polarization
vector of the photon. \hspace*{0.5cm}Expression (2) is valid for
an arbitrary external EM field. In the following we shall use it
for the case, namely conversions in the periodic EM field of the
resonant cavity. Here we use the following notations: $q \equiv
|\vec{q} |, p \equiv |\vec{p}|=(p_o^2-m_a^2)^{1/2}$
and $\theta$ is the angle between $\vec{p}$ and $\vec{q}$.\\

{\bf 3. Conversions in a resonant cavity}\\
\hspace*{0.5cm} For the sake of simplicity we choose the lowest
nontrivial solution of the resonant cavity, namely the $TM_{110}$
mode ~\cite{jac,lam}
\begin{eqnarray}
E_z &=& E_o\sin\left(\frac{\pi x}{a}\right)\sin \left(\frac{\pi
y}{b}
\right),\nonumber \\
H_x &=& \frac{i \varepsilon \pi}{\omega  b}E_o\sin\left(\frac{\pi
x}{a}\right)
\cos\left(\frac{\pi y}{b}\right),\nonumber\\
H_y &=& -\frac{i\varepsilon \pi}{\omega a}E_o\cos\left(\frac{\pi
x}{a}\right) \sin\left(\frac{\pi y}{b}\right), \label{tp}
\end{eqnarray}
 here the propagation of the EM wave is in the z - direction.
 Note that there is no E-wave of types $(000), (001), (010),
 (100), (101)$, and $(011)$.\\
\hspace*{0.5cm}Substitution of (\ref{tp}) to  $(2)$  gives us the
following expression for the matrix element
\begin{equation}
\langle p|{\cal M} |q\rangle =
\frac{g_{a\gamma}}{(2\pi)^{2}\sqrt{p_0 q_0}}
\left[\left(\varepsilon_1(\vec{q}, \tau)q_2 -
\varepsilon_2(\vec{q}, \tau)q_1\right)F_z + \varepsilon_1(\vec{q},
\tau)q_0F_x + \varepsilon_2(\vec{q}, \tau)q_0 F_y\right],
\end{equation}
where $ p_0\equiv q_0 + \omega$, and
\begin{eqnarray}
F_z &=& -\frac{8 E_0(q_x-p_x)(q_y-p_y)\cos[\frac{1}{2}a(q_x-p_x)]
\cos[\frac{1}{2}b(q_y-p_y)]\sin[\frac{1}{2}d(q_z-p_z)]}
{[(q_x-p_x)^2-\frac{\pi^2}{a^2}][(q_y-p_y)^2-\frac{\pi^2}{b^2}](q_z-
p_z)},\nonumber\\
F_x &=& -\frac{\varepsilon \pi^2 F_z}{\omega b^2(q_y - p_y)},\nonumber\\
F_y &=& \frac{ \varepsilon \pi^2 F_z}{\omega a^2(q_x - p_x)},
\end{eqnarray}
where $a$, $b$, and $d$ are three dimensions of the cavity and
$\omega$ is the frequency of the EM field. Substituting Eq.(5)
into Eq.(4) we finally obtain the DCS for conversions \\
\begin{eqnarray}
\frac{d\sigma(\gamma \rightarrow a)}
{d\Omega}&=&\frac{g^2_{a\gamma}p_o}{2(2\pi)^2 q_0}\left[({q_x}^2 +
{q_y}^2){F_z}^2 + (1 - \frac{{q_x}^2}{q^2}){q_0}^2 {F_x}^2 +
(1 - \frac{{q_y}^2}{q^2}){q_0}^2 {F_y}^{2}\right. \nonumber \\
& &\left. -2q_xq_y{F_x}{F_y} - 2{q_0}{q_x}{F_y}{F_z} +
2{q_0}{q_y}{F_x}{F_z}\right]. \label{dc1}
\end{eqnarray}
\hspace*{1cm}Assuming that the momentum of the photon is parallel
to the z - axis (the direction of the EM field) then Eq.
(\ref{dc1}) gets the final form
\begin{eqnarray}
\frac{d\sigma(\gamma \rightarrow a)} {d\Omega'}&=&\frac{8\pi^2
g^2_{a\gamma}E_0^2q^2}{\omega^2} (1 +
\frac{\omega}{q})\left[\frac{(p\sin\theta\cos\varphi')^2}{b^4 } +
\frac{(p\sin\theta\sin\varphi')^2}{a^4 }\right]\nonumber \\
& &\times\left[\frac{\cos\frac{a}{2}(p\sin\theta\cos\varphi')
\cos\frac{b}{2}(p\sin\theta\sin\varphi') \sin\frac{d}{2}(q -
p\cos\theta)} {[(p\sin\theta\cos\varphi')^2-\frac{\pi^2}{a^2})]
[(p\sin\theta\sin\varphi')^2-\frac{\pi^2}{b^2}](q-p\cos\theta)}\right]^2,
\label{dc2}
\end{eqnarray}
where $\varphi'$ is the angle between the x - axis and the
projection of ${\vec p}$ on the xy - plane and $d\Omega' =
d\varphi' d\cos\theta$ . From (7) it is easy to show that there
are no conversions when $\theta = 0$ and $\varphi' =
\frac{\pi}{2}$ and
\begin{eqnarray}
\frac{d\sigma(\gamma \rightarrow a)}{d\Omega'} & =&\frac{8
g^2_{a\gamma}E_0^2p^2}{\omega^2\pi^2\left(p^2 -
\frac{\pi^2}{a^2}\right)^2 }\left(1 + \frac{\omega}{q}\right)
 \cos^2(\frac{p a}{2})\sin^2(\frac{q d}{2}),
\label{dc3}
\end{eqnarray}
for $\theta =\frac{\pi}{2}$ and $\varphi' = 0$.\\
In the limit $q\rightarrow\frac{\pi}{d}$, Eq.(8) becomes
\begin{eqnarray}
\frac{d\sigma(\gamma \rightarrow a)}{d\Omega'}& = &\frac{8
g^2_{a\gamma}E_0^2p^2}{\omega^2\pi^2\left(p^2 -
\frac{\pi^2}{a^2}\right)^2 }\left(1 + \frac{\omega d}{\pi}\right)
 \cos^2(\frac{p a}{2}).
\label{dc3}
\end{eqnarray}
\hspace*{1cm}Next, if the momentum of the photon is perpendicular
to the direction of the  EM field,i.e., in the y - axis then from
Eq. (\ref{dc1}) we have
\begin{eqnarray}
\frac{d\sigma(\gamma \rightarrow a)} {d\Omega''}&=&\frac{32
g^2_{a\gamma}E_0^2q^2}{(2\pi)^2} (1 + \frac{\omega}{q})\left[(q
-p\cos\theta) - \frac{\pi^2}{\omega b^2 }\right]^2
\tan^2\varphi"\nonumber \\
& &\times\left[\frac{\cos\frac{a}{2}(p\sin\theta\sin\varphi")
\cos\frac{b}{2}
(q-p\cos\theta)\sin\frac{d}{2}p(\sin\theta\cos\varphi")}
{[(p\sin\theta\sin\varphi")^2-\frac{\pi^2}{a^2})]
[(q-p\cos\theta)^2-\frac{\pi^2}{b^2}]}\right]^2, \label{dc2}
\end{eqnarray}
where $\varphi''$ is the angle between the z - axis and the
projection of ${\vec p}$ on the xz - plane.\\
From (\ref{dc2}) we have $\frac{d\sigma(\gamma \rightarrow a)}
{d\Omega''} = 0$ for  $\theta = \varphi'' = 0$ and
\begin{eqnarray}
\frac{d\sigma(\gamma \rightarrow a)}{d\Omega''}& =
&\frac{g^2_{a\gamma}E_0^2 a^2d^2q^2\left( \omega q - \frac{\pi^2}{
b^2}\right)^2}{2(2\pi)^2 \omega^2 \left(q^2 -
\frac{\pi^2}{b^2}\right)^2 }\left(1 + \frac{\omega}{q}\right)
 \cos^2(\frac{q b}{2}),
\label{dc3}
\end{eqnarray}
for $\theta =  \varphi'' = \frac{\pi}{2}$, and in the limit
$p\rightarrow\frac{\pi}{a}$.
\begin{figure}
\centerline{\epsfxsize=3.2in\epsffile{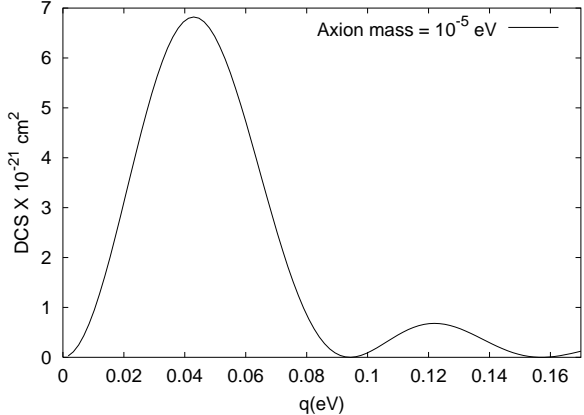}} \caption{DCS
as a function of the photon momentum $q$. Axion mass is taken to
be $m_a=10^{-5}$ eV.}
\end{figure}

\hspace*{1cm} From Eq.(11) we can see that this is the best case
for the conversions. In this case, the DCS depends quadratically
on the amplitude $E_0$, photon momentum $q$ and two dimensions of
the resonant cavity. \\
{\bf 4. Conclusion}\\
\hspace*{1cm}The following consequences may be obtained from our
results:\\
i) To compare the results with the wave guide( for details, see
Ref.~\cite{dvsoa}) we can introduce the DCS's ratio $R$ between
the $TM_{110}$ mode of the resonant cavity and  the $TE_{10}$ mode
of the wave guide ( for the best case)
\begin{eqnarray}
 R &= &\frac{(DCS)_{110}}{(DCS)_{10}}\sim \frac{q^2}{\omega^2}\left(1
+ \frac{\omega}{q}\right)^2. \label{dc3}
\end{eqnarray}
From Eq.(12) it is easy to see that, in the limit $\omega^2 =
m^2_a\ll q^2$ then we have $(DCS)_{110}\gg (DCS)_{10}$. This means
that {\it the conversion cross sections of the resonant cavity are
much larger than those of the wave guide in the same conditions}.\\
ii) We estimate the result for the best case in C. G. S. Gauss
units. Using data as in Refs.~\cite{sik,dvsoa}, $ E_o = 10^6
cm^{-1/2} g^{1/2}s^{-1}$, $ a = b = d = 100$ cm and $\omega = m_a$
, the DCS for Eq.(11) as a function of the momentum of photon was
plotted in the figure 1. We can see from the figure, when the
momentum of photons is perpendicular to the momentum of axions
then there exists {\it the resonant conversion } at the value $q
\approx 4.7\times 10^{-2}$ eV, the DCS is given by
$\frac{d\sigma(\gamma \rightarrow a)}{d\Omega'}$=
$6.7\times10^{-21} cm^2$. This is due to the limit
$q\rightarrow\frac{\pi}{b}$. At the high values of the momentum
$q$ then DCS's have very small values.\\
iii) We emphasize that the explicit resonance happens {\it only}
in the direction {\it perpendicular} to the direction of the
photon while in the static EM fields the axions are produced
mainly in the direction
of the photon motion.\\
\hspace*{1cm}Finally, it is worth noting that the scattering of
photons in an external EM field is the most important effect
because it has the following advantages~\cite{soa}: {\it Firstly},
this is the unique effect giving nonzero cross-section in the
first order of the perturbation theory. {\it Secondly}, since the
EM field is classical we can increase the scattering probability
as much as possible by increasing the intensity of the field or
the volume containing the field. {\it This is a important point in
order to apply it in experiments}. However in our work we
considered only a theoretical basis for experiments, other
problems with axion detection will be investigated in the
future.\\
\hspace*{1cm}This work was supported in part by the Vietnam
National Basic Research Programme
on National Sciences.\\

\end{document}